\documentclass[aps,prl,twocolumn,groupedaddress,showpacs]{revtex4}
\usepackage{graphicx}
\usepackage{bm}
\usepackage{hyperref}
\usepackage{dcolumn}% Align table columns on decimal point
\newif\ifbib
\bibfalse

\begin{document}

\title{Roton-Rotation Coupling of Acetylene in $^4$He}

\author{R.~E.~Zillich and K.~B.~Whaley}

\affiliation{Department of Chemistry, University of California,
Berkeley, CA 94720}

\begin{abstract}
Rotational absorption spectra of acetylene in superfluid $^4$He are
calculated using a combined correlated basis function-diffusion Monte
Carlo method (CBF/DMC) as well as a numerically exact path integral correlation
function approach (PICF).  Both approaches predict anomalously large
distortion constants in addition to reduced rotational constants, 
and the higher rotational excitations 
are seen to be accompanied by an additional peak and absorption band.
The CBF analysis shows that these phenomena are due to strong coupling of the 
higher rotational states with the roton and maxon excitations of $^4$He, and
the assocated divergence of the helium density of states in this region.
\end{abstract}

\pacs{05.30.Jp, 33.20.Bx, 33.20Sn, 36.40.-c, 67.40.Yv}

\maketitle

Helium nanodroplets have become a useful matrix for isolation spectroscopy
\cite{toennies01}.
Unlike in classical liquids, the spectra of embedded molecules may be 
rotationally resolved in the superfluid environment of $^4$He, 
which has led to numerous investigations
of rotational constants of molecules in $^4$He droplets (see table~1 in
Ref.~\cite{callegari2001JCP}). For all but the lightest rotors,
a significant reduction of the rotational constants with respect to their
gas phase values has been found, with identical symmetry,
which has been explained by the adiabatic following of
a molecular-interaction induced non-superfluid fraction of the solvation 
shell.\cite{focusarticle} 
In contrast, the lightest
molecules exhibit relatively small reduction of their
rotational constants in $^4$He droplets~\cite{callegari2001JCP}.  
Calculations have shown that 
the assumption of adiabatic following breaks down for light 
molecules~\cite{patel02} and that instead of coupling to localized
helium solvation density, there is appreciable coupling to delocalized 
collective helium modes~\cite{zillich04}. 

Acetylene (HCCH) is one such light molecule, whose ro-vibrational spectrum has 
been obtained by infra-red spectroscopy in $^4$He droplets~\cite{nauta01JCP}.
The effective rotational constant $B_{\rm eff}$ was found to be reduced to
88\% of its gas phase value (1.1766 cm$^{-1}$)
and a very large effective distortion constant $D_{\rm eff}$ was  measured, 
showing a relative enhancement $D_{\rm eff}/B$ an order of
magnitude larger than that seen for heavier molecules such as OCS, SF$_6$ and
(HCN)$_2$.
It was speculated that the excessive broadening of the IR spectral lines
might indicate coupling to excitations of the helium, 
but in the absence of theoretical understanding of such a coupling
no conclusions could be drawn.
The calculations of the rotational absorption spectra of HCCH
presented here show how coupling between the molecular
rotation and the roton excitations of $^4$He results in an
anomalously large distortion constant as well as a reduced rotational
constant, with predicted values in good agreement with experimental 
measurements. They also reveal that the higher rotational excitations
display a secondary peak as well as a weak maxon-roton band between them
that both result from this coupling.  We discuss
how these additional spectral features might be observed, in order to
stimulate experiments to measure this unique signature of rotation-roton
coupling. 

We present two independent calculations.
The first calculation is a combination of correlated basis function (CBF)
theory and diffusion Monte Carlo (DMC) methodology that 
derives zero temperature excitation energies from linear response theory
for correlated wave functions~\cite{SKJLTP} and then evaluates these with 
numerically exact DMC calculations for the required input
ground state properties~\cite{zillich04}. The rotational excitations
are described in this CBF/DMC approach within an analytic approximation that
reveals the microscopic origins of the rotational excitations and
associated spectral absorption features.  For the second calculation we
use path integral Monte Carlo methods~\cite{huang02,cuiPRB97} 
(at temperature $T=0.625$~K)
to evaluate the imaginary-time orientational correlation functions
$F_J(t) = (1/Z)\sum_m{\rm Tr}
\{Y_{Jm}^{+}(\Omega(t)) Y_{Jm}(\Omega(0))e^{-H/{kT}}\}$
($\Omega$ is the molecular orientation), sampled from 
the full molecule-helium Hamiltonian $H$ that includes molecular rotation and
all translational
%%BW went back to original statement of symmetry plus ref with note
degrees of freedom as well as the permutation exchange symmetry of
helium~\cite{note_c2h2perms}. The inverse Laplace transform 
(made with the maximum entropy method~\cite{bryan90})
of $F_J(t)$ is proportional to the rotational
absorption spectrum for dipole ($J=1$), quadrupole ($J=2$), and higher
order transitions. The PICF calculation provides a quantum simulation free from
systematic approximation which is important for 
verification of the novel spectroscopic features predicted by the 
CBF/DMC calculations.

In Ref.~\cite{zillich04}, we formulated the CBF/DMC approach for
molecular excitations in $^4$He and showed how CBF
theory is able to describe the
effect of coupling of collective excitations of the $^4$He environment
to the rotational dynamics of a molecule. This coupling was seen to
renormalize the rotational energy $\hbar\omega_J$ corresponding to total
angular quantum
number $J$ for a linear molecule with bare mass $M$ and gas phase 
rotational constant $B$
by an energy-dependent self energy $\Sigma_J(\omega_J)$ \cite{notesymm},
\begin{equation}
  \hbar\omega_J\ =\ BJ(J+1)\ +\ \Sigma_J(\omega_J)
\label{eq:omegaL}
\end{equation}
which is a functional of several ground state
quantities~\cite{zillich04}.
%Two of these, the static structure factor
%$S(p)$ and the phonon-roton dispersion $\epsilon(p)$ of bulk $^4$He
%are known experimentally~\cite{DonnellyDonnellyHills}.
The Legendre coefficients $g_{\ell'}(r)$, ${\ell'}>0$, of the pair distribution
function between HCCH and $^4$He are obtained from a DMC simulation
of HCCH and 256 $^4$He atoms (periodic boundary conditions), employing a
trial wave function of the form used in Ref.~\cite{zillich04} 
(parameters $c=7.724$\,\AA\ and $b=2.670$\,\AA),
interaction potentials from Refs.\cite{Aziz97,moszynskiJCP95},
and implementing descendent weighting~\cite{boronat95a}.
Odd components of $g_{\ell'}(r)$ vanish for HCCH and the zero component
$g_{\ell'=0}(r)$ does not contribute to $\Sigma_J(\omega)$, so that 
the $\ell'=2$ coefficient is the dominant contribution to $\Sigma_J(\omega)$.
From angular momentum conservation~\cite{zillich04},
one can then show that for HCCH the dominant effect of the superfluid
$^4$He environment on the molecular rotational excitations $J$
is the coupling of molecular rotational states that differ by $|J-\ell|=2$.

In  general, the self energy is complex, leading to finite life-times
and decay of rotational excitations into other modes. 
When the momentum $p$ is such that the input energy
$\hbar\omega$ equals the sum of three excitation energies
$B\ell(\ell+1)$ (molecular rotation), $\epsilon(p)$ (phonon-roton excitation),
and $\hbar^2p^2/2M$ (kinetic energy of the recoiling HCCH),
these states can be excited by decay of the rotational
excitation $\hbar\omega=\hbar\omega_J$.

Eq.~(\ref{eq:omegaL}) needs to be solved self-consistently for $\omega_J$
since the self energy $\Sigma_J(\omega_J)$ is itself a function of $\omega_J$. 
When the imaginary part of $\Sigma_J$ is small, the real part of the
solutions are seen to correspond to the positions of well-defined peaks in the 
dipole ($J=1$), quadrupole ($J=2$),\dots absorption spectra of the
molecule from its ground state. These spectra, normalized to unity, are given 
by 
\begin{eqnarray}
  S_J(\omega) &=& {1\over\pi}
        \Im m[BJ(J+1)\ +\ \Sigma_J(\omega) - \hbar\omega]^{-1}
  \nonumber\\
  &=& {\Im m\Sigma_J(\omega)/\pi\over
    (\gamma_J(\omega)-\hbar\omega)^2+(\Im m\Sigma_J(\omega))^2}\,,
\label{eq:SL2}
\end{eqnarray}
where we have defined $\gamma_J(\omega) = BJ(J+1)+\Re e\Sigma_J(\omega)$.
The imaginary part of the self energy, $\Im m\Sigma_J$,
gives the homogeneous width of the Lorentzian line associated with
decay of the excitation of life-time $\hbar /\Im m\Sigma_J(\omega_J)$
into the lower energy modes of the molecule (rotation and translation)
accompanied by emission of a $^4$He excitation.
In the opposite limit when the imaginary part of $\Sigma_J(\omega)$ is large,
eq.~(\ref{eq:SL2}) shows that the absorption $S_J(\omega)$ can become large 
regardless of whether or not $\omega$ is a solution of eq.~(\ref{eq:omegaL}).
In this case we find a broad band which cannot be uniquely associated with
a single molecular rotation mode, but instead is due
either to a simultaneous excitation of both the molecular rotation  and
$^4$He modes, or of only $^4$He modes (see below).

\begin{figure}%[h]
\centerline{
  \includegraphics[width=0.8\linewidth]{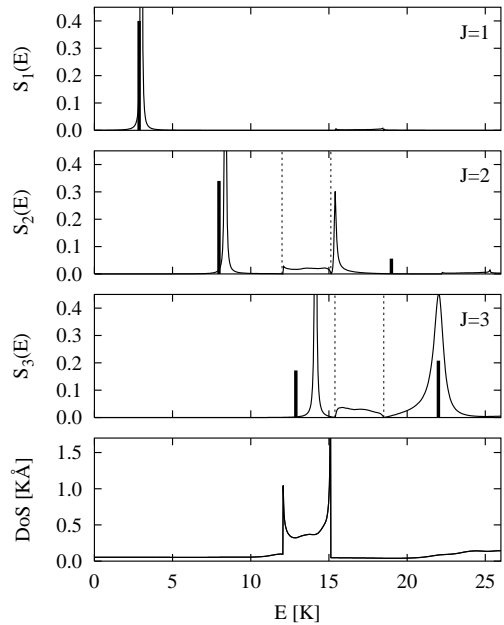}
}
\caption[]{\label{FIG:spectrum}
  CBF/DMC absorption spectra $S_J(\omega)$, $J=1,2,3$, for HCCH in 
  bulk $^4$He (thin solid lines) and PICF rotational
%BW 'height' rather than 'length'
  excitations (broad solid lines, height proportional to the spectral
  weight) for HCCH in a cluster of 64 $^4$He.
  The CBF/DMC spectra have been broadened by $10$~mK. 
  The bottom panel shows the density of states $\varepsilon'(p)+\hbar^2p/M$.
  The vertical dotted lines in the $J=2$ and $J=3$ panels
  indicate the edges of the 
  roton-maxon bands coupling these states to $\ell=0$ and $\ell=1$ states, 
  respectively.
}
\end{figure}

Fig.~\ref{FIG:spectrum} shows the ground state absorption spectra
of HCCH for $J=1,2,3$ calculated by CBF/DMC, eq.~(\ref{eq:SL2}).
The rotational excitations obtained from PICF are also shown, as broad
solid lines with height proportional to their spectral weight.
Lineshapes and broad spectral features cannot be obtained with the
correlation function spectral inversion approach, but
the positions and relative strength of discrete peaks are
very reliable~\cite{paesaniJCP03hydro}. The CBF/DMC spectra have sharp
Lorentzian peaks associated with excitations of finite life-time, as well as
broad features between these peaks.  The peaks 
are in good qualitative agreement with the PICF excitations,
which are made for HCCH in a cluster of 64 $^4$He atoms (approximately two
solvation layers). This implies that the relevant $^4$He collective 
excitations are already well developed at this cluster
size
and that the rotational coupling to these modes is not sensitive
to temperature variation below $T \sim 0.6$~K.

We analyze first the sharp peak structures.
The locations $\hbar\omega'$ of peaks in the CBF/DMC spectrum $S_J(\omega)$
are found by solving the equation $\gamma_J(\omega)-\hbar\omega = 0$ (see also
eq.~\ref{eq:SL2}), which can have one or more solutions.
Each solution $\hbar\omega'$ can be associated
with a rotational excitation of finite life-time, which can
decay into a $\ell<J$ state and an excitation of the helium environment.
It is evident from Fig.~\ref{FIG:spectrum} that while for $J=1$ a single
peak is obtained, the CBF/DMC absorption spectra
$S_J(\omega)$ for $J=2$ and $J=3$ have {\em two\/} peaks each,
in contrast to the expected pattern of excitations for a linear rotor which
has only a single excitation at energy
$BJ(J+1)$ for a given $J$. This surprising double peak feature
is also found for the rotational excitations calculated with the numerically
exact PICF method. The latter
do not rely on any analytic approximations and thus provide additional
verification of the correctness of the splitting seen in the CBF/DMC spectra.
We therefore are led to conclude that unlike heavier rotors such as OCS,
the pure rotational spectrum of HCCH in $^4$He for excitations higher than 
$J=1$ does not conform with the simple pattern predicted for an effective 
linear rotor.

Inspection of the CBF equations allows us to determine that the reason
for the occurrence of two peaks for absorptions to $J=2$ and $J=3$ is the
non-monotonic behavior of the collective
excitation spectrum in $^4$He.  The divergent density
of states (see bottom panel of Fig.~\ref{FIG:spectrum}) at both the
roton minimum (wave number $p_r$) and the maxon
maximum (wave number $p_m$) of the helium excitation
spectrum~\cite{DonnellyDonnellyHills} results in divergences in the self energy
$\Sigma_J(\omega)$ at $\omega=B\ell(\ell+1)+\varepsilon(p)+\hbar^2p^2/2M$ for
$p=p_r,p_m$ and $\ell=0,1,2,\dots$. 
These divergences can split spectral
peaks that are close to the roton and maxon into two,
shifting the ``primary'' peak below the roton
minimum and moving the ``secondary'' peak above the maxon
maximum~\cite{zillich04}.
The dipole spectrum, $J=1$, has only a
single peak, because the $J=1$ excitation energy
is much lower than the roton energy and consequently does not couple so
effectively to the collective modes.
Consequently the secondary $J=1$ absorption peak has
negligible weight and is not observed.
For $J=2$, the weight of the secondary peak is small but non-negligible,
while for $J=3$ more than half of the spectral weight
is actually carried by the secondary peak, see Table~\ref{tab1}.

\begin{figure}%[h]
\centerline{
  \includegraphics[width=0.9\linewidth]{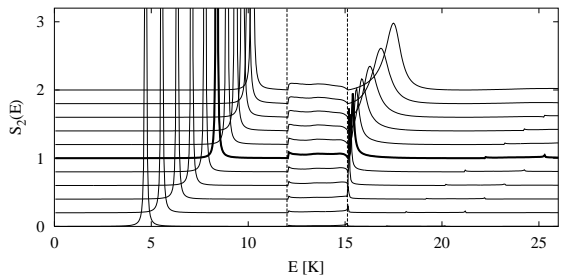}
}
\caption[]{\label{FIG:fac}
  The absorption spectrum $S_J(\omega)$, for the $J=0\to 2$ transition
  of HCCH in $^4$He, where we have varied the gas phase rotational constant
  $B$ between $0.5\times B$ (bottom curve) and $1.5\times B$ (top curve),
  in steps of $0.1\times B$.
}
\end{figure}
In order to further understand the coupling
between rotations and rotons that is implied by this splitting, we have 
calculated the CBF/DMC spectrum as a function of the molecular gas phase
rotational constant $B$.  This allows the molecular rotational energy to be
tuned across the roton/maxon energy regime, and the coupling between
molecule and helium to be modulated as a result.
Fig.~\ref{FIG:fac} shows the quadrupole spectrum $S_J(\omega)$,
$J=2$, as $B$ is artificially varied between
$0.5\times B$ and $1.5\times B$ in eq.~(\ref{eq:SL2}) and in the
expression for the self energy. 
For simplicity we use the same molecule-helium pair distribution function
for all $B$ values in this comparison. 
When the gas phase value of $B$ is sufficiently small that the $J=2$ 
excitation does not couple to the roton excitations it is evident that 
$S_2(\omega)$ now shows only a single peak, just like $S_1(\omega)$ in
Fig.~\ref{FIG:spectrum}.
On increasing $B$ to its actual gas phase value,
$1.1766$cm$^{-1}$, the secondary peak is seen to emerge and
simultaneously a weak roton-maxon band starts to
grow in between the two peaks. 
On increasing $B$ beyond the HCCH value, the secondary peak 
continues to grow until it eventually accounts for the entire spectral weight
and the primary peak vanishes.

Between these two peaks in the $J=2,3$ spectra, there are no sharp rotational
peaks because excitations here would decay instantaneously due to 
the high density of states between the roton and maxon excitations
of $^4$He. The resulting large imaginary part of $\Sigma_J(\omega)$
leads to a broad roton-maxon band in the spectrum.  (Unfortunately such 
broad features cannot be retrieved from analytic continuation
of the PICF to real frequencies.)
Although these roton-maxon bands appear small
in Fig.~\ref{FIG:spectrum}, they do possess non-negligible
weights of 0.06 and 0.08 in the $J=2$ and $J=3$ spectra, respectively.
The roton-maxon band for $J=3$ is
shifted in energy by $2B$ with respect to the corresponding band for $J=2$, 
because for $J=2$ the molecular rotation is coupled to a pure excitation of
the helium environment with the molecule in
its rotational ground state $\ell=0$, while for $J=3$, the rotation
is coupled to an $\ell=1$ molecular state.
In both cases $J-\ell=2$, which corresponds to the
strongest contribution to the self energy $\Sigma_J(\omega)$, as noted above.

We now consider how these CBF and PICF results may be observed experimentally.
Our ground state spectra correspond to dipole, quadrupole, etc. 
rotational absorption from the $J=0$ {\em ground state}, while
in Ref.~\cite{nauta01JCP} rotational transitions
$J=0 \leftrightarrow 1$ and $J=1 \rightarrow 2$ were observed (accompanied
by excitation of the C-H vibrational stretches).
\begin{table}%[H]
\centering
\renewcommand{\arraystretch}{1.25}
\begin{tabular}{c|cc|cc|c}
\hline
& \multicolumn{2}{c|}{CBF} & \multicolumn{2}{c|}{PICF} & Exp. \\
$J$ &  E [cm$^{-1}$] & $w$ & E [cm$^{-1}$] & $w$ & E [cm$^{-1}$] \\
\hline
\hline
1  &  2.08$\pm$0.04 &\ 0.98\ &\  1.99$\pm$0.10  \ &\ 1.00\ &\  2.01 \\
\hline
2  &  5.82$\pm$0.24 &\ 0.82\ &\  5.54$\pm$0.10  \ &\ 0.85\ &\  5.55 \\
2' & 10.70$\pm$1.31 &\ 0.08\ &\  13.20$\pm$0.90  \ &\ 0.14\ &\ --- \\
\hline
3  &  9.82$\pm$0.69 &\ 0.33\ &\  8.95$\pm$0.41  \ &\ 0.43\ &\  --- \\
3' & 15.28$\pm$0.70 &\ 0.49\ &\  15.28$\pm$0.53 \ &\ 0.52\ &\  --- \\
\hline
\end{tabular}
\caption{
  Rotational excitation energies of HCCH in $^4$He calculated with
  CBF (bulk $^4$He) and PICF ($^4$He$_{64}$),
  for $J=1,2,3$, compared with experimental values extracted from IR
spectra~\cite{nauta01JCP}.
  $J$' denotes the secondary peaks and $w$ the
  respective spectral weights.
\label{tab1}
}
\end{table}
Table~\ref{tab1} compares the CBF and PICF rotational
excitations
with the experimental values extracted in~\cite{nauta01JCP}. 
The two excitations $J=1$ 
and $J=2$ that were accessed
by the spectroscopic measurements are seen to be in excellent agreement with 
the theoretical values for $J=1$ and for
the primary peak of the $J=2$ spectrum.
(The CBF self energy and accuracy of results can be further systematically 
improved~\cite{zillich04}.) 
Table~\ref{tab2} compares the spectroscopic constants
$B_{\rm eff}$ and $D_{\rm eff}$ obtained by
fitting these two excitations to the non-rigid linear rotor spectrum
$B_{\rm eff}J(J+1)-D_{\rm eff}(J(J+1))^2$. The distortion
constant relative to $B$, $D_{\rm eff}/B$, is very large compared to that for
heavier molecules~\cite{callegari2001JCP,nauta01JCP}.  
Our calculations show that this anomalously large
distortion constant results from the strong coupling of the $J=2$ excitation
with the helium roton and maxon excitations.  Fig.~\ref{FIG:fac} shows that the
spectral shift to lower frequencies that is responsible for $D_{\rm eff}$ is 
greater for lighter molecules, reflecting an increased coupling as
the molecular excitation become more resonant with these helium modes that
derives primarily from their high density of states.  
In Ref.~\cite{zillich04}
we demonstrated that this deviation
from a linear rotor spectrum is much smaller when the molecule
couples to a simple linear (phonon) excitation spectrum.

\begin{table}%[H]
\centering
\renewcommand{\arraystretch}{1.25}
\begin{tabular}{c|ccc|c}
\hline
HCCH &\  CBF \ &\ PICF \ &\ Exp.\cite{nauta01JCP}\ &\ gas phase\cite{nauta01JCP} 
\\
\hline
\hline
$B_{\rm eff}$ &\ 1.075\ &\ 1.031\ &\ 1.042\ &\ 1.172\\
$D_{\rm eff}$ &\ 0.0175\ &\ 0.0179\ &\ 0.0195\ &\ 1.62$\times$10$^{-6}$\\
\hline
\end{tabular}
\caption{
Rotation constant $B_{\rm eff}$ and 
distortion constant $D_{\rm eff}$ of HCCH in $^4$He.
The CBF and PICF results are compared with the fit made to 
%%BW small rewording to be more precise and better distinguish difference
$J=0\rightarrow 1, J=1\rightarrow 2$ rovibrational fine structure transitions 
in IR spectra~\cite{nauta01JCP}. (Units in cm$^{-1}$)
\label{tab2}
}
\end{table}

In this CBF calculation the homogeneous linewidth of
the absorption spectra in Figs.~\ref{FIG:spectrum} and~\ref{FIG:fac}, 
$\Im m\Sigma_J(\omega)$, results from rotational
relaxation only. For $J=1$, the rotational linewidth vanishes
due to the symmetry of the HCCH-$^4$He potential (see eq.~(4.4) in
Ref.~\cite{zillich04}). For $J=2$, it is finite but still
smaller than the experimental IR linewidth, which is believed to also
have contributions from vibrational relaxation as well as inhomogeneous
broadening due to finite size effects~\cite{nauta01JCP}.
Vibrational relaxation
may lead to a reduced life-time of the
secondary transition such that it is too short to be observed.  Multi-phonon 
excitations, which become prevalent in $^4$He for excitation
energies around 20\,K and higher, and are not incorporated in the present CBF
analysis, may also broaden the theoretical peaks.  
%Estimation of the effect of these based on modification of 
%the self-energy denominator 
%to be consistent with Ref.~\cite{apaja} shows that incorporation of
%these will preferentially
%broaden the secondary, higher energy, peak, although not to the extent that it
%disappears.  
For HCCH the secondary peak for 
$J=2$ is predicted to lie outside the range accessed in Ref.~\cite{nauta01JCP}
and a preliminary search for this at higher IR energies has not been
successful~\cite{GaryPrivatecomm}, possibly for the above reasons. 
Another complication for comparison with the IR spectra is that 
the $J=2$ state is accessed by the dipole transition $J=1 \rightarrow 2$, 
while the CBF calculation yields the quadrupole transition
%%zil ``transition matrix elements'' is not good
$J=0 \rightarrow 2$, with different transition matrix elements for
the HCCH-$^4$He coupling. Measurement of the rotational Raman spectrum may
allow a direct access to the $J=0\to 2$ transition.

In summary, calculations of the rotational absorption spectra corresponding to 
dipole, quadrupole, etc. transitions of HCCH from its rotational ground 
state in bulk $^4$He, $0\to J$, using the CBF/DMC and PICF approaches
yield an anomalously large distortion constant $D_{\rm eff}$ that is seen to 
result from a strong coupling of the higher molecular rotation states to
the roton excitation of helium. The CBF/DMC spectra for $J=2$ and $J=3$ show
secondary peaks at higher energies, which are confirmed by the numerically
exact PICF results, as well as weak roton-maxon bands between the two peaks.  
We expect that these unique
signatures of coupling between roton modes of helium and rotational modes of 
embedded molecules will be found in Raman spectra, as well as 
possibly in microwave and IR spectra for other light molecules. 

\begin{acknowledgments}

We thank Ad van der Avoird for providing the
HCCH-He potential routine, Gary E.~Douberly and
Roger E.~Miller for fruitful discussion,
and the Central Information Services of the Kepler University in
Linz, Austria, for computational resources.
This work was supported by the Miller Institute for Basic Research in Science
(R.E.Z.) and by the NSF under grant CHE-010754.

\end{acknowledgments}

\ifbib
\bibliography {my,ocshehy,cluster2}

\begin{thebibliography}{17}
\bibitem[{\citenamefont{note_c2h2perms}(2004)\citenamefont{note_c2h2perms}}]{note_c2h2perms}
\bibinfo{note}{
We neglect nuclear symmetry of HCCH since at $T=0.625$K
HCCH is primarily in the rotational ground state.
}
\bibitem[{\citenamefont{notesymm}(2004)\citenamefont{notesymm}}]{notesymm}
\bibinfo{note}{
Application to the rotational structure of a vibrational transition
$0\to\{\nu_i\}$ requires that the argument of the CBF correlation
function include also the vibrational coordinates, and results in a
vibrational offset of the pure rotation spectrum. Imposition of the
symmetry of HCCH then restricts the sum over $\ell$ in Eq.~(2.34)
of Ref.~\cite{zillich04} to even (odd) values for even (odd) vibrational
transitions.
}
\end{thebibliography}
\else
%%%%%%%%%%%%%%%%%%%%%%%%%%%%%%%%%%%%%%%%%%%%%%%%%%%%%%%%%%%%%%%%%%%%%%%%%%%%%%%

%%%%%%%%%%%%%%%%%%%%%%%%%%%%%%%%%%%%%%%%%%%%%%%%%%%%%%%%%%%%%%%%%%%%%%%%%%%%%%%
\fi

\end{document}